\documentclass[submission,copyright,creativecommons]{eptcs}
\providecommand{\event}{GasCom 2024} 

\usepackage{underscore}         
\usepackage[T1]{fontenc}        

\usepackage{amssymb,amsmath,graphicx,ucs,hyperref}

\newtheorem{conjecture}{Conjecture}

\title{Square-Triangle Tilings:\\Lift \& Flip to Sample?}
\author{Thomas Fernique
\institute{HSE\\ Moscow, Russia}
\email{fernique@hse.ru}
\and
Olga Mikhailovna Sizova
\institute{Semenov Institute of Chemical Physics\\ Moscow, Russia}
\email{olstet@mail.ru}
}
\def\titlerunning{Square-triangle tilings}
\def\authorrunning{Th. Fernique, O. M Sizova}

\begin{document}
\maketitle

\begin{abstract}
We introduce an elementary transformation called flips on tilings by squares and triangles and conjecture that it connects any two tilings of the same region of the Euclidean plane.
\end{abstract}

\section{Introduction}

\begin{figure}[hbt]
\includegraphics[width=\textwidth]{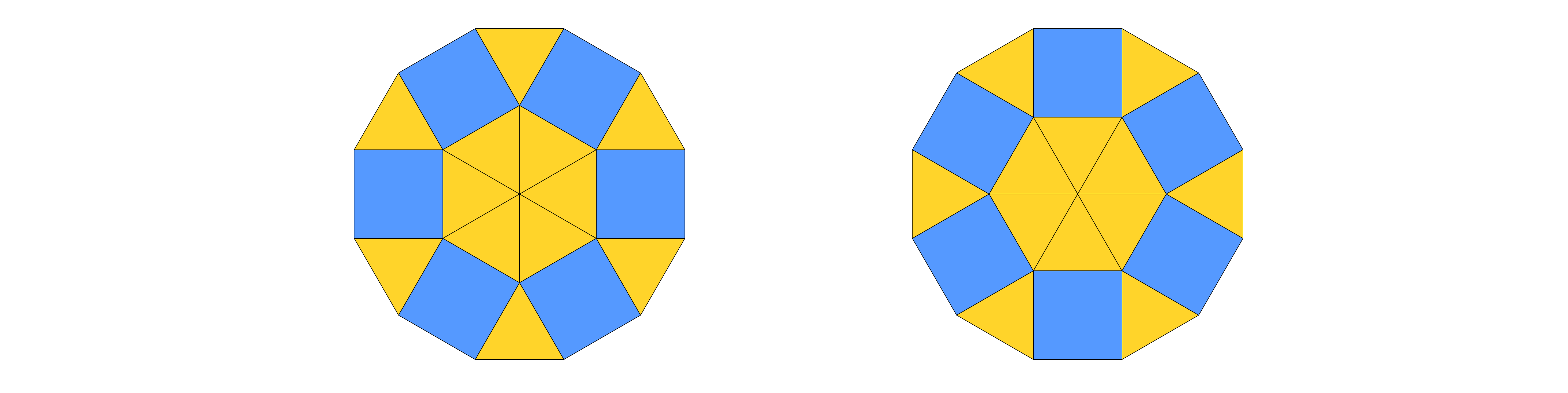}
\caption{Two square-triangle tilings of the same region.}
\label{fig:tilings}
\end{figure}

We consider the tilings of a simply-connected bounded region of the Euclidean plane by two tiles: a unit square and a regular unit triangle (Fig.~\ref{fig:tilings}).

The goal is to introduce an elementary local transformation that allows to travel the space of all the possible tilings, in order to sample them using mixing times techniques as in, e.g., \cite{LRS01}.
To do this, we shall embed these tilings in a larger set obtained by adding a rhombus tile, and show how these tilings can be naturally seen as discrete surfaces in 4-dimensional Euclidean space.
This is a work-in-progress.

\section{Lift}

Let $j=\exp\frac{2i\pi}{3}$ and $a=\exp\frac{2i\pi}{12}$ be the third and twelfth roots of the unity, respectively.
For $k=0,1,2$, define the complex numbers $u_k:=j^k$ and $v_k=a\cdot j^k$, seen as vectors in the Euclidean plane.
Without loss of generality, the edges of the square and triangles tiles are directed by the $u_i$'s and $v_i$'s.

\begin{figure}[hbt]
\centering
\includegraphics[width=\textwidth]{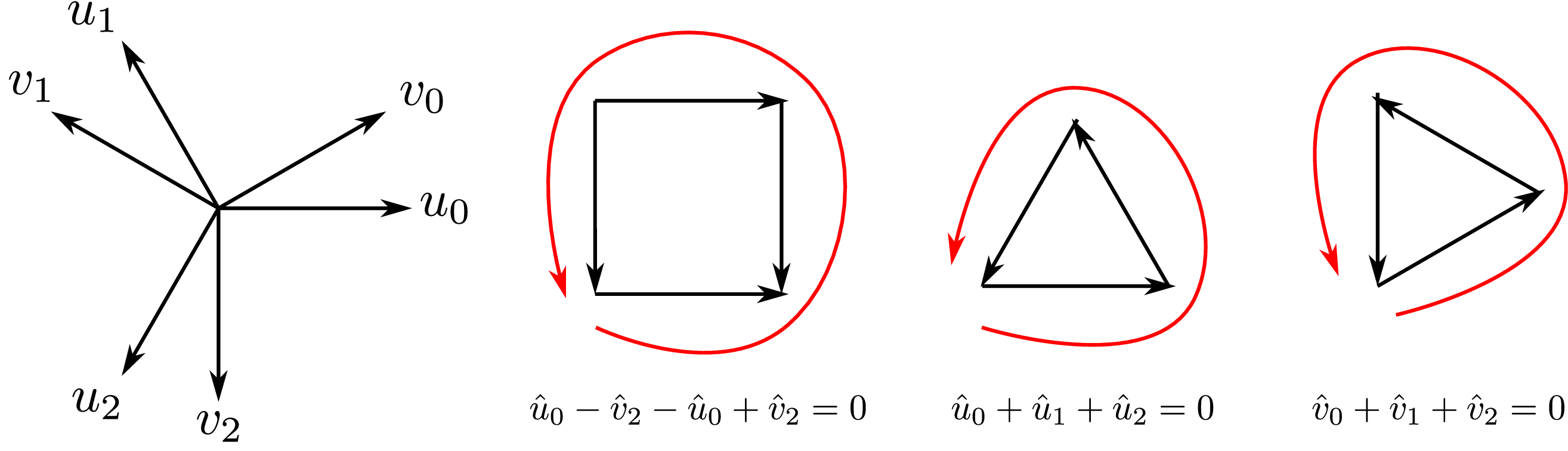}
\caption{The vectors which define tiles (left) and the conditions on their lifts.}
\label{fig:lift}
\end{figure}

We want to associate with every edge $w_k\in\{u_k,v_k\}$ a vector $\hat{w}_k$ in some higher dimensional space $\mathbb{R}^n$ so that, when travelling around a tile, the sum of the vectors associated with the traversed edges (with negative sign if travelled in backward direction) is zero.
This does not yield any restriction for the squares since each pair of parallel edges are travelled in opposite directions, hence their sum is always zero.
The same holds for any parallelogram (this will be used later).
For the triangles, this yields two conditions (Fig.~\ref{fig:lift}):
$$
\hat{u}_0+\hat{u}_1+\hat{u}_2=0
\quad\textrm{and}\qquad
\hat{v}_0+\hat{v}_1+\hat{v}_2=0.
$$

Since we have $2$ constraints on $6$ vectors, that is, a linear system of rank $4$, it is natural to take vectors in $\mathbb{R}^4$ (higher dimension is useless, lower dimension will loose generality).
We will define them as vectors in $\mathbb{C}^2$.
For example, take
$$
\forall i\in\{0,1,2\},\qquad
\hat{u}_k:=(j^k,0)
\qquad\textrm{and}\qquad
\hat{v}_k:=(0,j^k).
$$

\begin{figure}[hbt]
\centering
\includegraphics[width=\textwidth]{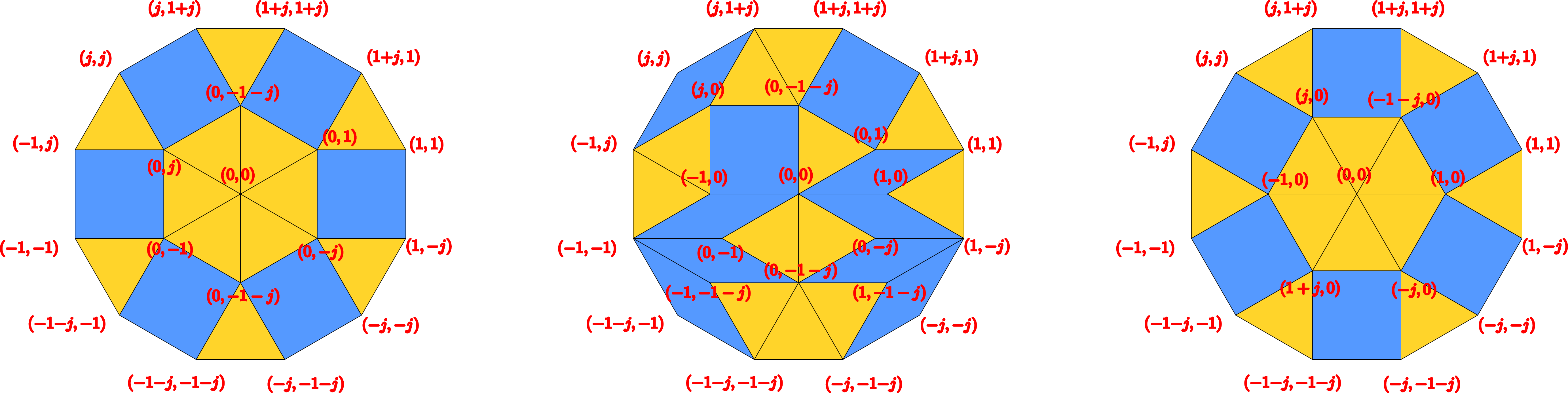}
\caption{Square-triangle tilings with the height of every vertex indicated.}
\label{fig:lift2}
\end{figure}

The condition on the sum of vectors around each tile ensures (by induction) that the sum of the vectors associated with the traversed edges along every cycle is zero.
In particular, if we map an arbitrary vertex $s_0$ of the tiling to $0\in\mathbb{C}^2$, then the sum $h(s)\in\mathbb{C}^2$ of vectors associated with the edges along a path from $v_0$ to a vertex $s$ of the tiling does not depend on the path.
The vector $h(s)$ is called the {\em height} of $s$.
This allows to see any square-triangle tiling as a sort of ``stepped'' two-dimensional surface embedded in $\mathbb{C}^2$.
Figure~\ref{fig:lift2} illustrates this.

Every triangle has edges directed either by three $u_k$'s or by three $v_k$'s.
We call it a {\em $u$-triangle} in the former case, a {\em $v$-triangle} in the latter case.
The way the $\hat{u}_k$'s have been chosen ensure that, in a lifted tiling, the vertices of every $u$-triangle have all the same second coordinate (in $\mathbb{C}$); this common coordinate defines the {\em elevation} of the triangle.
These three vertices differ by their first coordinate (in $\mathbb{C}$).
The barycenter of these coordinates is called the {\em position} of the triangle.
The situation is similar for $v$-triangles, with first and second coordinates exchanged.

\section{Flip}

\begin{figure}[hbt]
\centering
\includegraphics[width=0.3\textwidth]{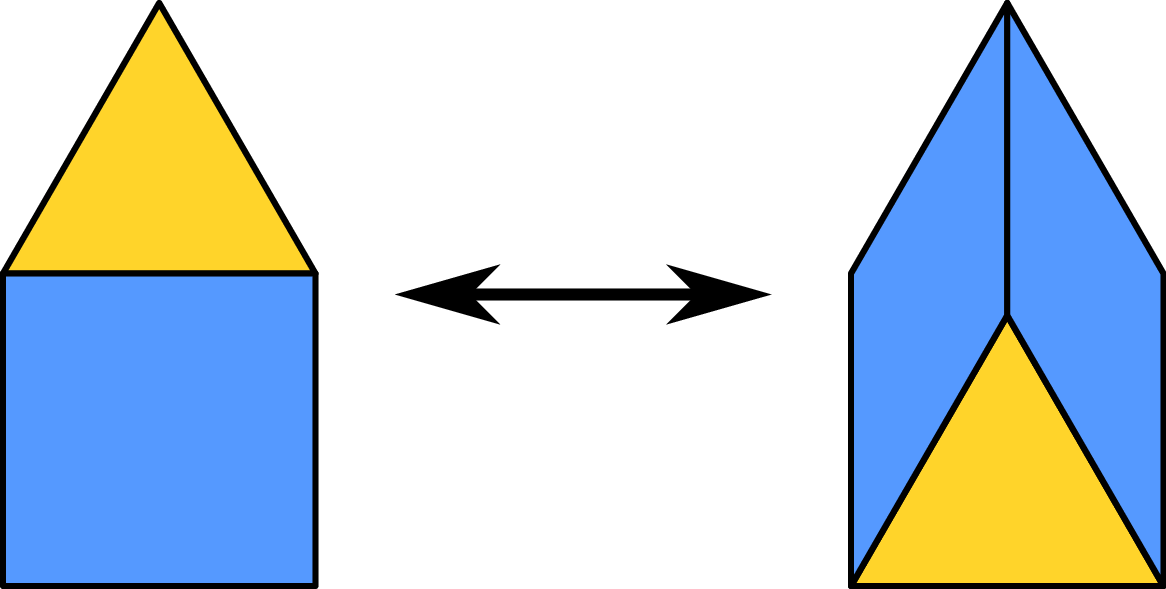}
\caption{A flip.}
\label{fig:flip}
\end{figure}

In the context of tilings, it is common to introduce elementary transformations on tilings generally called {\em flips}.
For example, in the case of tilings by rhombi, a flip denotes the half-turn rotation of a hexagon formed by three rhombi.
Square-triangle tilings lack of such a ``natural'' flip.
However, such a local transformation has been defined in \cite{OH93}, which features a third tile, namely a rhombus.
This flip is depicted in Figure~\ref{fig:flip}.

\begin{figure}[hbt]
\centering
\includegraphics[width=0.55\textwidth]{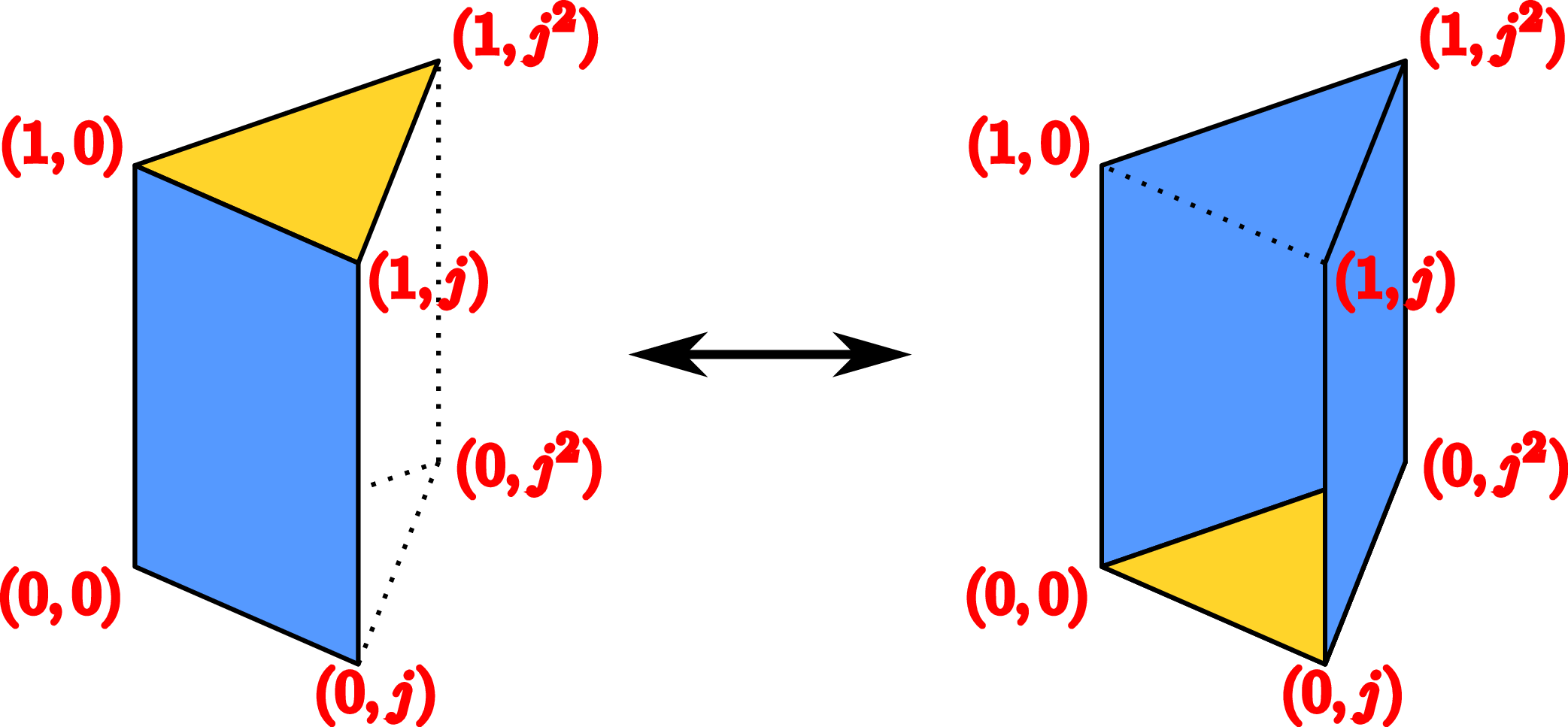}
\caption{A flip seen in the lift.}
\label{fig:fliplifted}
\end{figure}

It may seem strange to involve a new tile that takes us out of the space of square-triangle tilings.
But if we look on this flip in the lifted tiling\footnote{As for squares, pairs of parallel edges in rhombi do not yield any restriction for the lift.}, it can be seen as an exchange between the faces of a triangular prism: the upper triangle and one ``square'' lateral face are replaced by the lower triangle and the other two ``square'' lateral faces (Figure~\ref{fig:fliplifted}).
We are arguing here that squares and rhombi play the same role, differing only in the way they are represented in the Euclidean plane.
To emphasize this, we use the same color for both tiles.
We will also still use the term square-triangle tilings even when rhombi are present, with the special cases of lack of rhombi being called {\em pure} square-triangle tilings.
The position and elevation of a triangle are affected as follows by a flip:
\begin{itemize}
\item the position does not change;
\item the elevation changes by $\pm j^k$ for some $k\in\{0,1,2\}$.
\end{itemize}

\section{Flip-connectivity}

In \cite{OH93} is defined the {\em zipper move}, which is a sequence of flips between two pure square-triangle tilings (while intermediary tilings may not be pure).

Here, we would like to prove that, any two square-triangle tilings of the same region (not necessarily pure, that is, with possibly rhombus tiles) can be connected by a sequence of flips.
This would yield the above claim as a corollary.

\begin{figure}[hbt]
\centering
\includegraphics[width=0.85\textwidth]{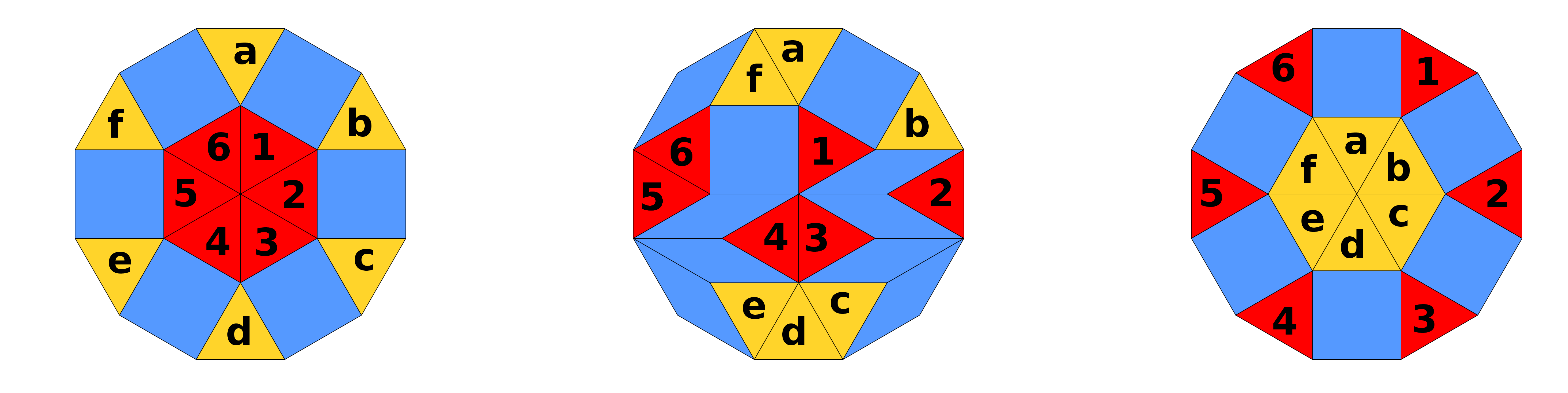}
\caption{Every pair of triangles with the same position have the same label.}
\label{fig:spottedtriangles}
\end{figure}

Recall that the position of a triangle is unchanged by a flip, only its elevation.
Hence, if flip-connectivity holds, then for any pair $T$ and $T'$ of tilings of the same region, to each triangle of $T$ corresponds a triangle of $T'$ with the same position.
Transforming $T$ into $T'$ by flips thus amounts to equal the elevations of pairs of triangles with identical position.
In other words, we can ``spot'' a triangle in every possible tilings.
Figure~\ref{fig:spottedtriangles} illustrates this.

Hence, the wanted result will follows if we can prove the following result (a similar result have been proven for rhombus tilings in \cite{BFR08}):

\begin{conjecture}
If $T\neq T'$ are two tilings of the same region, then there is a triangle in $T$ that can be moved by a flip towards its elevation in $T'$, so that no triangle which has already at the same elevation in $T$ and $T'$ is moved.
\end{conjecture}

\bibliographystyle{eptcs}
\bibliography{squaretriangle}
\end{document}